%% file: Paper-1085.tex
\documentclass[runningheads]{llncs}
\usepackage[T1]{fontenc}
\usepackage{booktabs}
\usepackage{adjustbox}
\usepackage{multirow}
\usepackage{pifont}
\usepackage{mathtools}
\usepackage{bbm}
\usepackage[ruled,vlined,linesnumbered]{algorithm2e}
\usepackage{pifont}
\usepackage{bbm}
\usepackage{amsmath}
\usepackage{amsfonts}
\usepackage{amssymb}
\usepackage{hyperref}
\usepackage{cleveref}
\usepackage{caption}
\usepackage{marvosym}
\usepackage{color, url}

\DeclarePairedDelimiter\floor{\lfloor}{\rfloor}
\newcommand{\cmark}{\textcolor{green}{\ding{51}}}%
\newcommand{\xmark}{\textcolor{red}{\ding{55}}}%
\Crefname{figure}{Fig.}{Figs.}
\Crefname{equation}{Eq.}{Eqs.}
\Crefformat{equation}{#2Eq.~#1#3}

\makeatletter
\newcommand{\printfnsymbol}[1]{%
  \textsuperscript{\@fnsymbol{#1}}%
}
\makeatother

\newcommand{\change}[1]{{\color{black}#1}} % blue

\begin{document}
\title{Heteroscedastic Uncertainty Estimation Framework for Unsupervised Registration}
\titlerunning{Heteroscedastic Uncertainty for Unsupervised Registration}
\author{Xiaoran Zhang\thanks{Equal contribution.}\inst{1}$^{(\text{\Letter})}$%\orcidID{0000-1111-2222-3333} 
\and Daniel H. Pak\printfnsymbol{1}\inst{1}%\orcidID{1111-2222-3333-4444} 
\and Shawn S. Ahn\inst{1,7} \and\\%\orcidID{2222--3333-4444-5555}
Xiaoxiao Li\inst{3,4,5} 
\and Chenyu You\inst{6}
\and Lawrence H. Staib\inst{1,5} \and\\
Albert J. Sinusas\inst{1,5,8} 
\and Alex Wong\inst{2}
\and James S. Duncan\inst{1,5}
}
% index{Zhang, Xiaoran}
% index{Pak, Daniel}
% index{Ahn, Shawn}
% index{Li, Xiaoxiao}
% index{You, Chenyu}
% index{Staib, Lawrence}
% index{Sinusas, Albert}
% index{Wong, Alex}
% index{Duncan, James}

\authorrunning{X. Zhang et al.}
\institute{Biomedical Engineering, Yale University, New Haven, USA\\
\email{xiaoran.zhang@yale.edu}
\and
Computer Science, Yale University, New Haven, USA
\and
Department of Electrical and Computer Engineering, The University of British Columbia, Vancouver, Canada
\and
Vector Institute, Toronto, Canada
\and 
Radiology \& Biomedical Imaging, Yale University, New Haven, USA
\and 
Electrical Engineering, Yale University, New Haven, USA
\and
Department of Surgery, University of Pennsylvania, Philadelphia, USA
\and
Department of Internal Medicine (Cardiology), Yale University, New Haven, USA
}
\maketitle              % typeset the header of the contribution

\input{sections/0_abstract}

\input{sections/1_intro}

\input{sections/2_related_works}

\input{sections/3_methods}

\input{sections/4_results}

\input{sections/5_conclusion}

\begin{credits}
\subsubsection{\ackname} This work is supported by NIH grant R01HL121226.

\subsubsection{\discintname}
The authors have no competing interests to declare that are
relevant to the content of this article. 
\end{credits}
\bibliographystyle{splncs04}
\bibliography{Paper-1085}

\input{sections/supplementary}
\end{document}

%% file: sections/0_abstract.tex
\begin{abstract}
%This paper proposes a heteroscedastic uncertainty estimation framework for unsupervised medical image registration. 
Deep learning methods for unsupervised registration often rely on objectives that assume a uniform noise level across the spatial domain (e.g. mean-squared error loss), but noise distributions are often heteroscedastic and input-dependent in real-world medical images. Thus, this assumption often leads to degradation in registration performance, mainly due to the undesired influence of noise-induced outliers. To mitigate this, we propose a framework for heteroscedastic image uncertainty estimation that can adaptively reduce the influence of regions with high uncertainty during unsupervised registration. The framework consists of a collaborative training strategy for the displacement and variance estimators, and a novel image fidelity weighting scheme utilizing signal-to-noise ratios. Our approach prevents the model from being driven away by spurious gradients caused by the simplified homoscedastic assumption, leading to more accurate displacement estimation. To illustrate its versatility and effectiveness, we tested our framework on two representative registration architectures across three medical image datasets. Our method consistently outperforms baselines and produces sensible uncertainty estimates. The code is publicly available at \url{https://voldemort108x.github.io/hetero_uncertainty/}.
% \keywords{First keyword  \and Second keyword \and Another keyword.}
\end{abstract}

% an adaptive weighting scheme with pixelwise relative $\gamma$-exponentiated signal-to-noise ratios (SNR) generated by a variance estimator

%% file: sections/1_intro.tex
\section{Introduction}
\label{sec:intro}
Deformable image registration solves for a dense pixel-wise displacement map that aligns one image to another. It is a key medical image analysis technique that enables disease progression monitoring and surgical guidance \cite{hill_medical_nodate,oliveira_medical_nodate,ta2024multi,zhang2021fully}. Classical methods (e.g. elastic-type models \cite{klein_elastix_2010}, free-form deformation with b-splines \cite{rueckert_nonrigid_1999}, diffeomorphic models \cite{ashburner_fast_2007}) are often computationally expensive and impractical for large-scale analyses. Thus, numerous works have proposed training a neural network to accelerate the test-time prediction of the displacement map \cite{balakrishnan_voxelmorph_2019,chen_transmorph_2022,hoffmann_synthmorph_2022,hoopes_hypermorph_2021,zhang_learning_2022,zhang2020comparative}. Such frameworks commonly utilize an unsupervised objective by minimizing the mean-square error (MSE) between the warped and target images. By using this objective, the existing frameworks assume an additive \textit{homoscedastic} Gaussian image noise, i.e. $\epsilon \sim \mathcal{N}(0, \sigma^2)$, with a constant noise variance across the spatial domain. However, this assumption is problematic for medical images, where the noise is intrinsically \textit{heteroscedastic} and \textit{input-dependent} (e.g. MRI \cite{eklund_bayesian_2017,wegmann_bayesian_2017} or ultrasound \cite{ouzir_data-adaptive_2020,zhang_ultrasound_2021}). The non-uniform noise variance across the image space (\Cref{fig:main_method}, left) is due to multiple factors including changes in anatomical structures or patient motion artifacts. The simplified homoscedastic assumption disregards such variations in noise levels and results in undesired penalization of false noise-induced outliers, causing unnatural and inaccurate deformations.

% davatzikos_spatial_1997, avants_symmetric_2008
% dalca_unsupervised_2018, shi_xmorpher_2022
% nascimento_robust_2008, stanziola_sparse_2019

% \begin{figure}[tb]
%     \centering
%     % \includegraphics[scale=0.1]{figures/variance_intro.png}
%     \includegraphics[scale=0.15]{figures/motivation.png}
%     % \includegraphics[scale=0.2]{figures/variance.pdf}
%     % \fbox{\rule[-.5cm]{0cm}{4cm} \rule[-.5cm]{4cm}{0cm}}
%     \caption{Existing unsupervised registration frameworks utilize an objective such as mean-squared error that assumes homoscedastic noise across an image \cite{balakrishnan_voxelmorph_2019,chen_transmorph_2022}. This does not reflect the heteroscedastic and input-dependent characteristics of noise in real-world medical imaging data. To address this issue, we propose a heteroscedastic uncertainty estimation scheme to adaptively weigh the data-fidelity term accounting for the non-uniform variations of noise across the image.}
%     \label{fig:motivation}
% \end{figure}

%A number of works were proposed to model heteroscedastic uncertainty in imaging problems \cite{kendall_what_2017,seitzer_pitfalls_2022} but we argue that directly applying such frameworks to image registration will result in significant performance decline. This is due to that the heteroscedastic uncertainty estimation shares a distinct objective from image registration, and thus two modules cannot be optimized jointly. 

To mitigate this, we propose a probabilistic heteroscedastic noise modeling framework for unsupervised image registration (\Cref{fig:main_method}, right). Our method involves two modules: (1) a standard displacement estimator and (2) a variance estimator that predicts the input-specific heteroscedastic noise for each image pair. %To optimize our framework properly, we introduce a collaborative learning strategy that alternatingly optimizes a displacement estimator and variance estimator with separate objectives, avoiding the joint optimization issue mentioned above. 
For proper utilization of the predicted noise, we introduce a novel adaptive weighting strategy based on the relative $\gamma$-exponentiated signal-to-noise ratio (SNR). This enables the successful convergence of our collaborative training strategy, where the two estimators improve upon each other via information exchange and loss calibration. %$\gamma$ reflects the confidence level of the current variance estimation. 
We validated the effectiveness and versatility of our proposed framework with extensive experiments using two representative neural network architectures and three cardiac datasets. Our proposed framework can be operated in a plug-and-play manner, and provides sensible heteroscedastic uncertainty measures to reflect spatially varying noise. This work may help provide a new perspective in data-driven input noise modeling for deep learning-based unsupervised registration. %within each image pair. We also conducted paired t-tests to demonstrate our improvements are statistically significant.

% : 1) ACDC (public 2D MRI) \cite{bernard_deep_2018}, 2) CAMUS (public 2D ultrasound) \cite{leclerc_deep_2019}, and 3) a private 3D echocardiography dataset (results shown in supplementary). To demonstrate its versatility, we implemented our proposed framework using two representative registration architectures: 1) Voxelmorph \cite{balakrishnan_voxelmorph_2019} and 2) Transmorph (\cite{chen_transmorph_2022}).

\textbf{Our contributions} include (1) We first analyze a naive approach of applying heteroscedastic noise modeling to unsupervised registration and identify the pitfalls of such an approach. (2) From which, we propose a probabilistic framework for data-driven estimation of heteroscedastic uncertainty for unsupervised registration. (3) We introduce an adaptive $\gamma$-exponentiated relative SNR weighting strategy for proper loss calibration. (4) We demonstrate the effectiveness and versatility of our method on two representative registration architectures across three datasets, demonstrating consistent statistically significant improvements over the baselines while providing sensible uncertainty maps.
% \begin{itemize}
%     \item We propose a heteroscedastic uncertainty estimation framework for unsupervised registration that extends the previous homoscedastic assumption.
%     \item We introduce an adaptive $\gamma$-exponentiated relative signal-to-noise weighting for displacement estimator to improve registration performance under the collaborative learning strategy together with a separate variance estimator.
%     \item We demonstrate the effectiveness and versatility of our proposed \textit{plug-and-play} framework on two representative registration architectures across three datasets with consistent statistically significant improvements over baselines while providing accurate uncertainty maps.
% \end{itemize}

% summary of contribution
% 1) We propose a probabilistic framework for data-driven estimation of heteroscedastic uncertainty for unsupervised registration. 
% 2) We introduce an adaptive $\gamma$-exponentiated relative SNR weighting strategy for proper loss calibration. 
% 3) We demonstrate the effectiveness and versatility of our method on two representative registration architectures across three datasets, demonstrating consistent statistically significant improvements over the baselines while providing sensible uncertainty maps.

%% file: sections/2_related_works.tex
\section{Related works} \label{sec:related_works}

\subsection{Unsupervised image registration}
% \paragraph{\textbf{Unsupervised image registration.}}
Voxelmorph is a popular \change{convolutional} neural network (CNN) model for unsupervised registration that was trained with the MSE loss, which assumes homoscedastic input noise \cite{balakrishnan_voxelmorph_2019}. Extensions of this work include Voxelmorph-diff \cite{dalca_unsupervised_2019}, %with a probabilistic diffeomorphic architecture to model displacement uncertainty, 
Hypermorph \cite{hoopes_hypermorph_2021}, %with an amortized hyperparameter optimization scheme, 
and Synthmorph \cite{hoffmann_synthmorph_2022} with similar assumptions. 
%with a contrast-invariant design without acquired images. 
Transformers have also been explored for unsupervised \change{registration} by replacing the convolutional encoders in the Voxelmorph U-Net with swin transformer layers \cite{chen_transmorph_2022}. Transformers can increase the receptive field of the encoding branch, potentially improving the long-range modeling capability \cite{liu_swin_2021}.  Further extensions include replacing the convolutional decoders with transformer layers \cite{shi_xmorpher_2022} and introducing multi-scale pyramids \cite{ma_pivit_2023}.

In this paper, we selected Voxelmorph \cite{balakrishnan_voxelmorph_2019} and Transmorph \cite{chen_transmorph_2022} as the representative registration architectures and tested our proposed framework and other baselines on both across different datasets. We also compared Voxelmorph-diff \cite{dalca_unsupervised_2019} with a direct extension of our framework that incorporates the \textit{displacement} uncertainty, in addition to our main contribution of modeling \textit{image} uncertainty. \change{The detailed analysis can be found in the Supp. Mat.}
% as described in \cref{sec:exp_sigma_z}.

\subsection{Heteroscedastic uncertainty estimation}
% \paragraph{\textbf{Heteroscedastic uncertainty estimation.}}
To model the heteroscedastic uncertainty in imaging problems, Kendall et al.~\cite{kendall_what_2017} proposed a joint mean-and-variance optimization strategy by minimizing the negative log-likelihood (NLL) under the maximum likelihood estimation (MLE) framework. %via stochastic gradient descent. 
%The objective includes a data-fidelity term with inverse heteroscedastic variance weighting and a regularization term on the predictive variance. 
This formulation has been effective in many applications, such as surface normal estimation \cite{bae_estimating_2021} and image segmentation \cite{monteiro_stochastic_2020}. Seitzer et al. further advanced the formulation with $\beta$-NLL to address the undesired undersampling effects of the inverse variance weighting of the data fidelity term~\cite{seitzer_pitfalls_2022}. In this paper, we selected both NLL \cite{kendall_what_2017} and $\beta$-NLL \cite{seitzer_pitfalls_2022} as our baselines for heteroscedastic uncertainty estimation.
% by arguing that the inverse variance weighting on data fidelity results in undersampling, causing performance degradation. They propose a

\subsection{Adaptive weighting schemes}
% \paragraph{\textbf{Adaptive weighting schemes.}}
A wide range of imaging problems is cast into optimizing an energy function, which contains a data-fidelity term and a regularization term. The relative importance between the two terms is usually tuned via hyperparameter optimization. However, such a strategy disregards the heteroscedastic nature of the error residuals \cite{wong_adaptive_2021}. Several adaptive weighting schemes have been proposed to address this issue by adjusting the weighting throughout optimization \cite{hong_adaptive_2017,wong_bilateral_2019}. Wong et al.~\cite{wong_adaptive_2021} extended the formulation to multi-frame setting. In this paper, we selected AdaReg \cite{wong_bilateral_2019} and AdaFrame \cite{wong_adaptive_2021} as baselines to compare with our proposed adaptive signal-to-noise weighting scheme.

% , accounting for non-uniform characteristics of the errors

% hong_adaptive_2017-1

%% file: sections/3_methods.tex
\section{Analysis of a naive approach}
\label{sec:prelim}
% \revise{We first present the vanilla approach by incorporating heteroscedastic noise modeling into unsupervised registration.} 
Let $I: \Omega\rightarrow \mathbb{R}$ be an image \change{and $\Omega \subseteq \mathbb{R}^d$ its spatial domain}. Unsupervised image registration aims to find the alignment between a moving image $I_m$ and a fixed image $I_f$, \change{without} ground truth information about warping displacement field $z: \Omega\rightarrow\mathbb{R}^d$. %The estimated displacement $\hat{z}$ is obtained from a displacement estimator using a neural network $\hat{z}=q_\theta(I_m,I_f)$ with parameters $\theta$. 
For each image pair $(I_m,I_f)$, we assume $I_m$ is a fixed parameter and $I_f$ is a noisy observation of the warped image $\hat{I}_f = I_m \circ z$. Then, $z$ is a sample from the posterior distribution $p(z|I_f; I_m)$.
% where $\circ$ denotes image warping.

With this formulation, the main task is finding the displacement posterior $p(z|I_f; I_m)$ that will maximize the data likelihood $p(I_f; I_m)$. Since directly solving for $p(z|I_f; I_m)$ is intractable,  we apply the variational approach to approximate it using a neural network $q_{\theta}(z|I_f;I_m)$. To minimize the discrepancy between approximate and real posterior distributions, we compute the KL-divergence and arrive at the negative evidence lower bound:
\begin{align}
    \begin{split}
        &\text{KL}(q_{\theta}(z|I_f;I_m)\|p(z|I_f;I_m)) = \text{KL}(q_{\theta}(z|I_f;I_m)\|p(z)) \\
        &\qquad\qquad\qquad\qquad\;\;\; - \mathbb{E}_{q_\theta}(\log p(I_f|z;I_m)) + \text{const.} %\log p(I_f;I_m)
    \end{split}
    \label{eq:full_exp}
\end{align}

For the warped image distribution $p(I_f|z; I_m)$, we assume an additive \change{zero-mean} heteroscedastic Gaussian noise with spatially varying variance $\sigma_I^2:\Omega\rightarrow \mathbb{R}^+$ to incorporate input-dependent heteroscedasticity
\begin{align}
    p(I_f|z; I_m) = \mathcal{N}(I_f; \hat{I}_f, \sigma_I^2). \label{eq:sigma_I_noise}
\end{align}

Similar to \cite{balakrishnan_voxelmorph_2019,kendall_what_2017}, we assume pixel-wise independence, isotropic Gaussian for the displacement posterior $q_{\theta}(z|I_f; I_m) = \mathcal{N}(z; \mu_{z}, \sigma_{z}^2\mathbb{I})$, and uniform Gaussian for the displacement prior $p(z) = \mathcal{N}(0, \mathbb{I})$. Then, we arrive at the preliminary loss to be minimized for each image pair
\begin{equation}
    \mathcal{L} = \mathbb{E}_\Omega \left[ \underbrace{\frac{1}{\sigma_{I}^2} [I_f-\hat{I}_f\|]^2}_{\mathcal{L}_{\text{data}}} + \log\sigma_{I}^2 + \lambda\|\nabla z\|^2 \right],
    \label{eq:prelim_loss_logIonly}
\end{equation}
where $\hat{I}_f = I_m \circ z$. $\sigma_I^2$ is estimated by our variance estimator as $\hat{\sigma}_I^2 = h_\phi(I_f, \hat{I}_f)$, \change{using} the fixed and moved image pair as input. $z$ is estimated by our displacement estimator as $\hat{z}=q_\theta(I_m,I_f)$, \change{using} the moving and fixed pair as input.

\section{Our approach}
\label{sec:methods}
% \paragraph{\textbf{Collaborative heteroscedastic uncertainty estimation.}} %\label{sec:collaborative_learning}
\subsection{Collaborative heteroscedastic uncertainty estimation}
\begin{figure*}[tb]
    \centering
    \begin{tabular}{c|c}
        \includegraphics[scale=0.11]{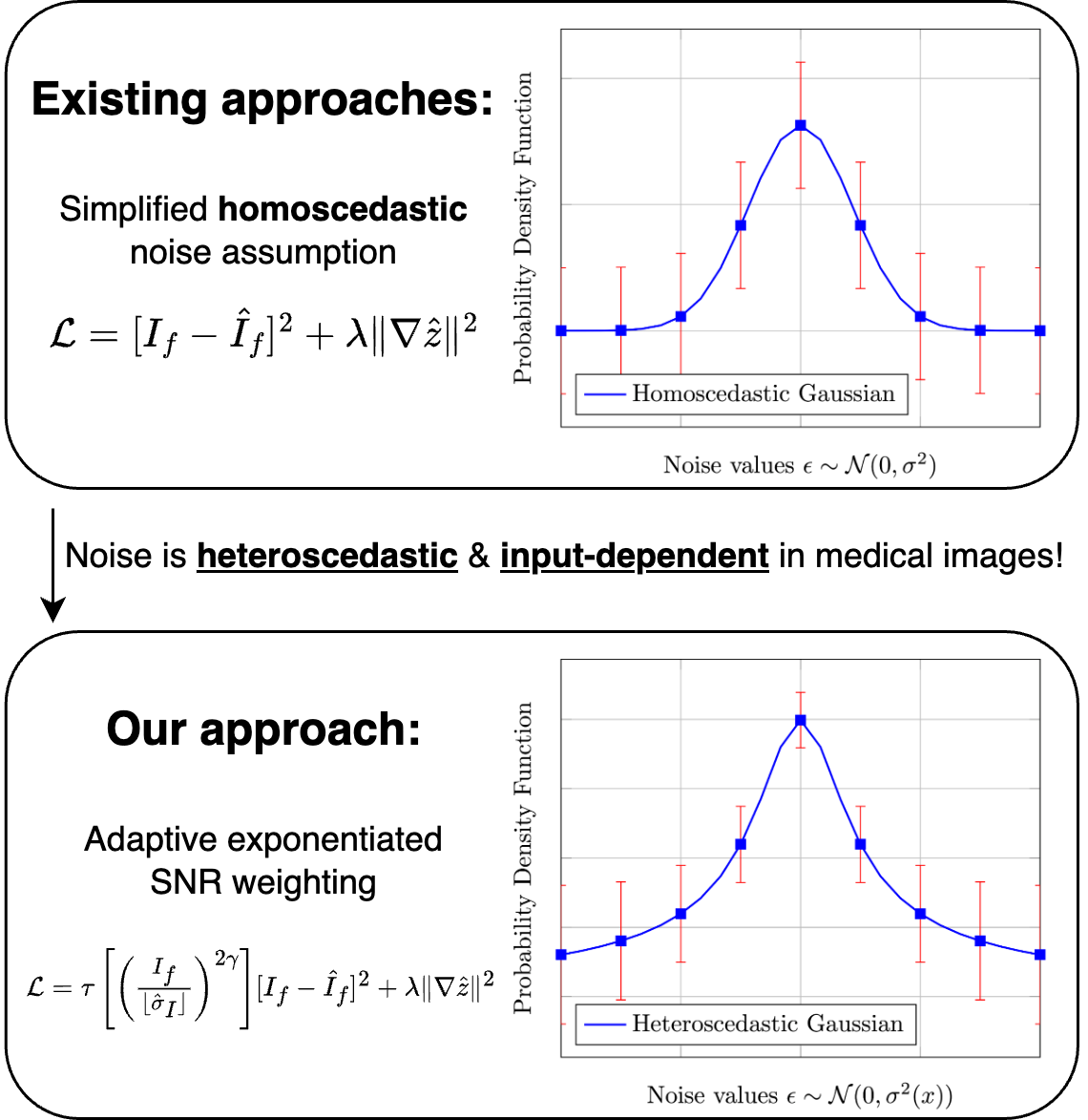} &  \includegraphics[scale=0.06]{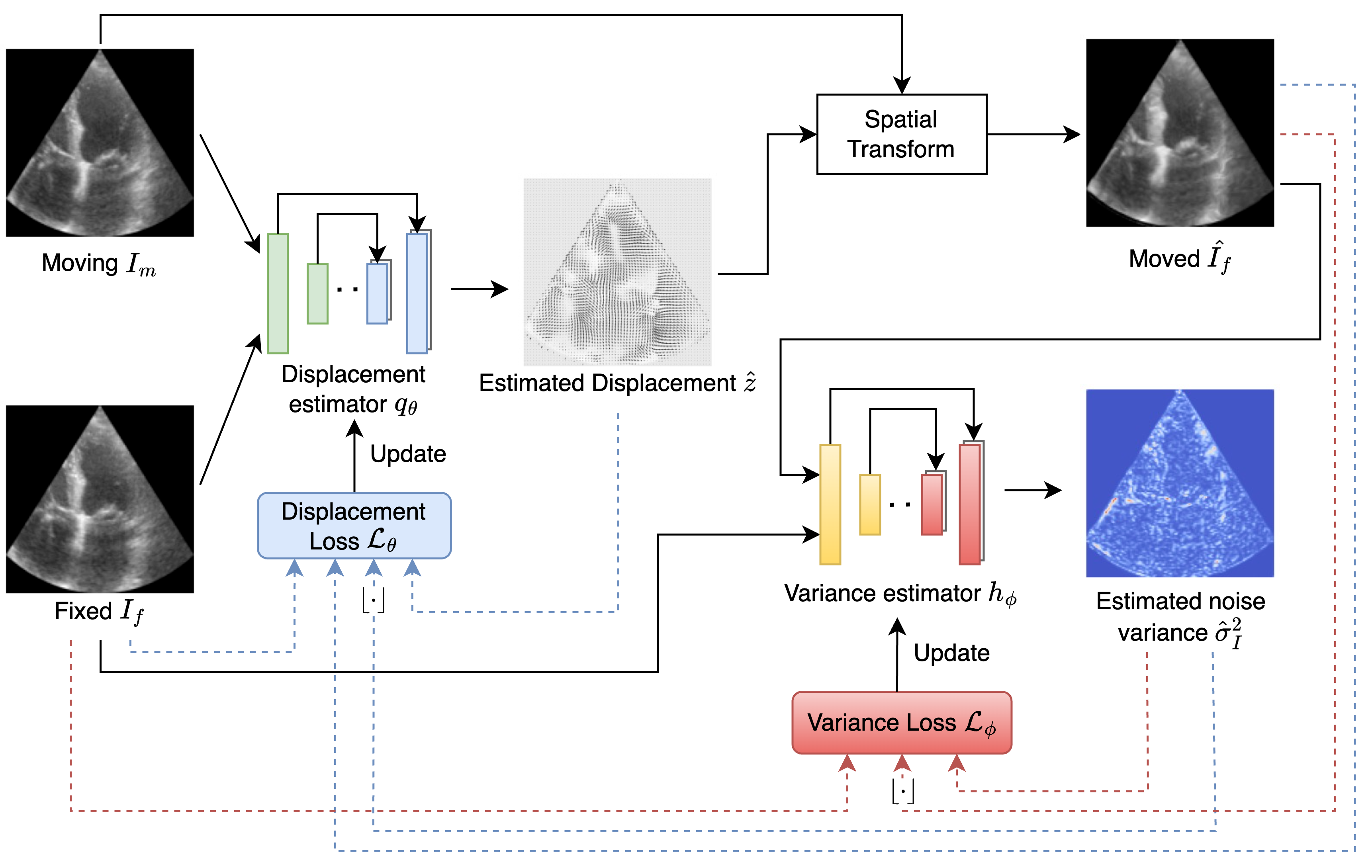}
    \end{tabular}
    \caption{Left: We propose a heteroscedastic uncertainty estimation scheme to adaptively weight the data-fidelity term accounting for the non-uniform variations of noise across the image. Right: Overview of our proposed method. %Our displacement estimator is built upon the U-Net backbone with final separate convolutional layers to predict the displacement. 
    The noise variance estimator uses a U-Net backbone that takes reconstructed frame $\hat{I}_f$ along with frame $I_f$ to predict the heteroscedastic variance for the noise in \Cref{eq:sigma_I_noise}.}
    \label{fig:main_method}
\end{figure*}

Thus far, the formulation implies a single objective (\Cref{eq:prelim_loss_logIonly}) for the optimization of both displacement and variance estimators. Even with perfectly estimated variance, such na\"ive application may result in undersampling of higher intensity regions \cite{seitzer_pitfalls_2022}, as their noise levels tend to be elevated due to the relatively preserved SNR throughout the image space. To mitigate this, we propose a learning strategy that utilizes separate objectives for the two modules, but with collaborative information exchange (\Cref{fig:main_method}). Specifically, the displacement estimator is optimized using \Cref{eq:displacement_loss}, which largely depends on the quality of the predicted variance. In parallel, the variance estimator directly uses the displacement estimator output (i.e. the warped image) as a part of its input to predict the variance.
% that utilizes \Cref{eq:prelim_loss_logIonly} only as the objective for the variance estimator, and separately defines the objective of the displacement estimator.
% \revise{We also validate the necessity by comparing the proposed approach with joint vanilla optimization strategies (NLL and $\beta$-NLL) on unsupervised registration in \Cref{sec:reg_acc}.}

% This effect is observed in the displacement estimator within the image fidelity term $\mathcal{L}_{\text{data}}$ due to the application of absolute inverse variance weighting, further resulting in performance degradation. 

% In our proposed formulation (\Cref{fig:main_method}), the displacement and the variance estimators work \textit{collaboratively} using different objectives to update the network parameters while sharing intermediate predictions (i.e. estimated displacement $\hat{u}$ and variance $\hat{\sigma}_I^2$) to mutually enhance their performance.

% which is also dependent on the quality of the reconstructed image (moved) warped using the estimated displacement

% \paragraph{\textbf{Adaptive weighting using exponentiated relative SNR.}} %\label{sec:loss_function}
\subsection{Adaptive weighting using exponentiated relative SNR}
Our proposed formulation above offers the flexibility of designing separate objectives for the displacement and variance estimators. For the displacement loss, we addressed the undersampling issue by modifying the image fidelity weighting term from the inverse of \textit{absolute} variance to the $\gamma$-exponentiated \textit{relative} SNR. Our proposed displacement loss is defined as
\begin{align}
    \mathcal{L}_{\theta} = \mathbb{E}_\Omega \left[ \mathcal{T}\left[ \left(\frac{I_f}{\floor{\hat{\sigma}_I}}\right)^{2\gamma}\right] \|I_f-\hat{I}_f\|_2^2 + \lambda \|\nabla \hat{z}\|^2 \right], \label{eq:displacement_loss}
\end{align}
where $\mathcal{T}(\cdot)$ is the sigmoid activation function. The hyperparameter $\gamma$ indicates the confidence of the displacement model on the estimated variance; when $\gamma=0$ the image fidelity term reduces to MSE. Gradient stopping (denoted as $\floor{\cdot}$) was necessary to treat the weighting term as a scalar map and to prevent duplicate back-propagation by the two estimators.

% , the image reconstruction error reduces to MSE, whereas when $\gamma=1$, the reconstruction error is weighted higher toward regions with less noise

% that encourages the network to concentrate on regions with relatively less noise. %reducing the negative impact of undersampling. 

% We experiment with different settings of $\gamma$ described in \Cref{sec:reg_acc} and discover that a value of $\gamma=0.5$ achieves the optimal trade-off.
% , showing that the displacement model is less confident in the variance estimate
% Through hyperparameter optimization, we set $\gamma=0.5$

% as our variance loss to update parameters $\phi$

For optimizing the variance estimator, we used the $\beta$-NLL objective \cite{seitzer_pitfalls_2022}
\begin{align}
    \mathcal{L}_{\phi}= \mathbb{E}_\Omega \left[ \floor{\hat{\sigma}_I^{2\beta}} \left( \frac{1}{\hat{\sigma}_I^2} \|I_f-\floor{\hat{I}_f}\|_2^2 +\log \hat{\sigma}_I^2 \right) \right].\label{eq:variance_loss}
\end{align}
Our network was implemented to output $\log \hat{\sigma}_I^2$ for numerical stability. We updated each estimator in an alternating fashion using two separate optimizers.

%% file: sections/4_results.tex
\section{Datasets and results}
\label{sec:results}
% \textbf{\textit{Datasets.}}
\subsection{Datasets}
We tested our method on three distinctive cardiac datasets: (1) \textbf{ACDC} \cite{bernard_deep_2018}: Human MRI, multiple 2D slices + time, 150 total sequences. (2) \textbf{CAMUS} \cite{leclerc_deep_2019}: Human echocardiography, 2D + time, 1000 total sequences from 500 patients, (3) \textbf{Private 3D Echo}: \textit{in vivo} Porcine, \textit{in vivo} canine, and synthetic echocardiography, 3D + time, 99 total sequences. All datasets included segmentations of the left ventricular myocardium at end-diastole (ED) and end-systole (ES). For all of our experiments, we warped the ED frame to reconstruct the ES frame. The training, validation, and testing splits were 60/20/20(\%). Images were preprocessed with resizing and normalization (Supp. Mat.). Implementation details (hyperparameters, compute, etc.) are in the Supp. Mat.

\subsection{Registration accuracy}
\label{sec:reg_acc}
% \paragraph{\textbf{Quantitative and qualitative evaluations.}}
\subsubsection{Quantitative and qualitative evaluations.}
\begin{figure*}[t]
    \centering
    \begin{tabular}{c}
         \includegraphics[scale=0.24]{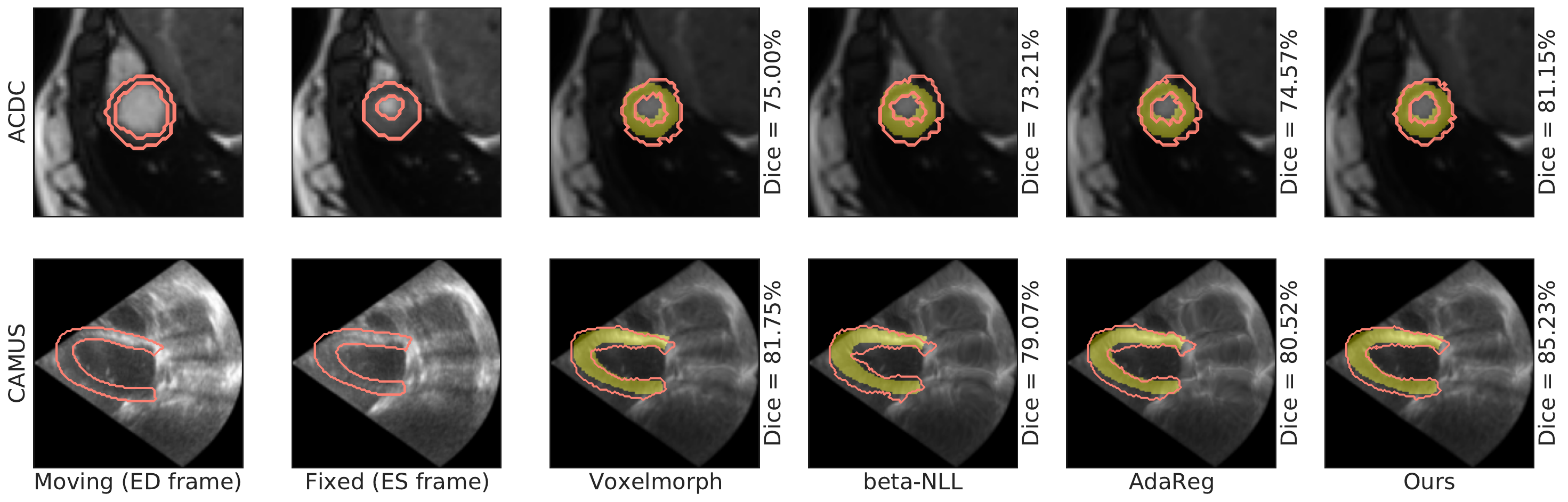} \\\midrule
         \includegraphics[scale=0.24]{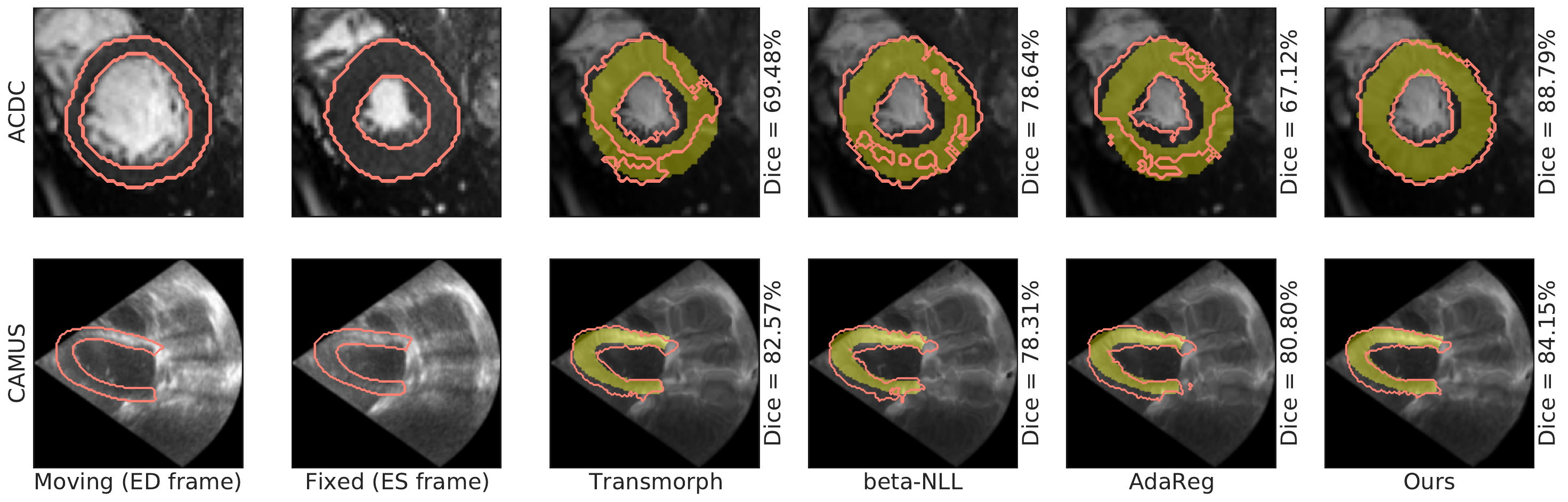}
    \end{tabular}    
    % \fbox{\rule[-.5cm]{0cm}{4cm} \rule[-.5cm]{4cm}{0cm}}
    \caption{Qualitative evaluation of the registration accuracy via segmentation warping for all datasets (top two rows: Voxelmorph architecture \cite{balakrishnan_voxelmorph_2019}, bottom two rows: Transmorph architecture \cite{chen_transmorph_2022}). Our method in the last column (overlayed with ground truth (GT) ES myocardium label in yellow) predicts more natural and accurate deformations compared to baselines, evidenced by better matching with the GT, smoother contour edges, and locally consistent myocardial region.}
    % Rows: different datasets. Colums: test-set labels or different models' warped segmentation.
    \label{fig:contour_main}
\end{figure*}

We present extensive quantitative evaluations of our method performance, where we show a consistent improvement from both classical and deep learning baselines across all datasets (\Cref{tab:contour}, \ref{tab:contour_echo}) with statistical significance (\Cref{tab:p_values}). Qualitative visualizations also demonstrate improved registration performance of our approach with smoother contour edges and more locally consistent myocardium (\Cref{fig:contour_main}).

% These results demonstrate the effectiveness and versatility of our proposed adaptive weighting based on relative SNR using our predicted variance. 

% \input{tables/contour_table}
\input{tables/combined_table}

% \input{tables/p_values}

% \paragraph{Comparison with vanilla joint optimization approaches}
Our consistent improvements from the vanilla version of Voxelmorph (vxm) \cite{balakrishnan_voxelmorph_2019} and Transmorph (tsm) \cite{chen_transmorph_2022}, which both utilize homoscedastic noise assumptions, especially confirm the benefit of heteroscedastic noise modeling (\Cref{tab:contour}).
% showing that not capturing spatially varying noise as in the homoscedastic objective results in less accurate registration. 

We note that both NLL \cite{kendall_what_2017} and $\beta$-NLL failed to improve upon the vanilla baselines (\Cref{tab:contour}). This validates our analysis that the joint training of displacement and variance estimators would degrade registration performance due to the undersampling from inverse absolute variance weighting (\Cref{sec:methods}). This also shows the advantage of our proposed collaborative learning framework, which provides the flexibility of designing separate objectives for the two estimators. Furthermore, we observe that existing adaptive weighting schemes of AdaReg \cite{wong_bilateral_2019} and AdaFrame \cite{wong_adaptive_2021} are ineffective, potentially due to the their assumptions on error residuals failing to represent the complicated real-world data distribution; this is in contrast to our proposed data-driven SNR-based weighting scheme. For 3D echo data, we show quantitative results in \Cref{tab:contour_echo} and qualitative results in the Supp. Mat. Note: Improvements come solely from smoother optimization and does not incur additional complexity during inference.

% \paragraph{\textbf{Effect of $\gamma$.}}
\subsubsection{Effect of $\gamma$.}
The exponentiated hyperparameter $\gamma$ provides the flexibility to adjust the estimated variance's degree of influence on the displacement estimator during training. This can be viewed as a trade-off in the predicted uncertainty estimates, where $\gamma=0$ means no confidence, reducing the adaptive weighting map to a uniform scalar map, and $\gamma=1$ indicating full confidence. \change{We empirically chose $\gamma=0.5$ for the best performance across all datasets (\Cref{tab:gamma}).} %Considering the overall performance across all datasets, our method performed best at $\gamma=0.5$ (\Cref{tab:gamma}).

%%%%%%%%%%%%%%%%%%%%%%%%%%%
% briefly mention this
%%%%%%%%%%%%%%%%%%%%%%%%%%%
% To validate the robustness of our method against other noise assumptions, we performed the same evaluations with a modified Eq. \ref{eq:variance_loss} that reflects a heteroscedastic Laplacian image noise. The overall performance improvements were essentially replicated for all three datasets, albeit with a slightly lower overall performance for all methods compared to the Gaussian image noise assumption.

% we first ranked all pixels based on their estimated uncertainty magnitudes
% then plotted the MSE of the remaining pixels ($\mathbb{E}_{\Omega-\Omega_x}[I_f-\hat{I}_f]^2$) versus the percentage of the removed pixels $\Omega_x$ based on the ranking

\subsection{Evaluation on heteroscedastic uncertainty} \label{sec:exp_sigma_I}
\begin{figure}[tb]
    \centering
    \begin{tabular}{c|c}
      \multirow{2}{*}[1.75cm]{\includegraphics[scale=0.17]{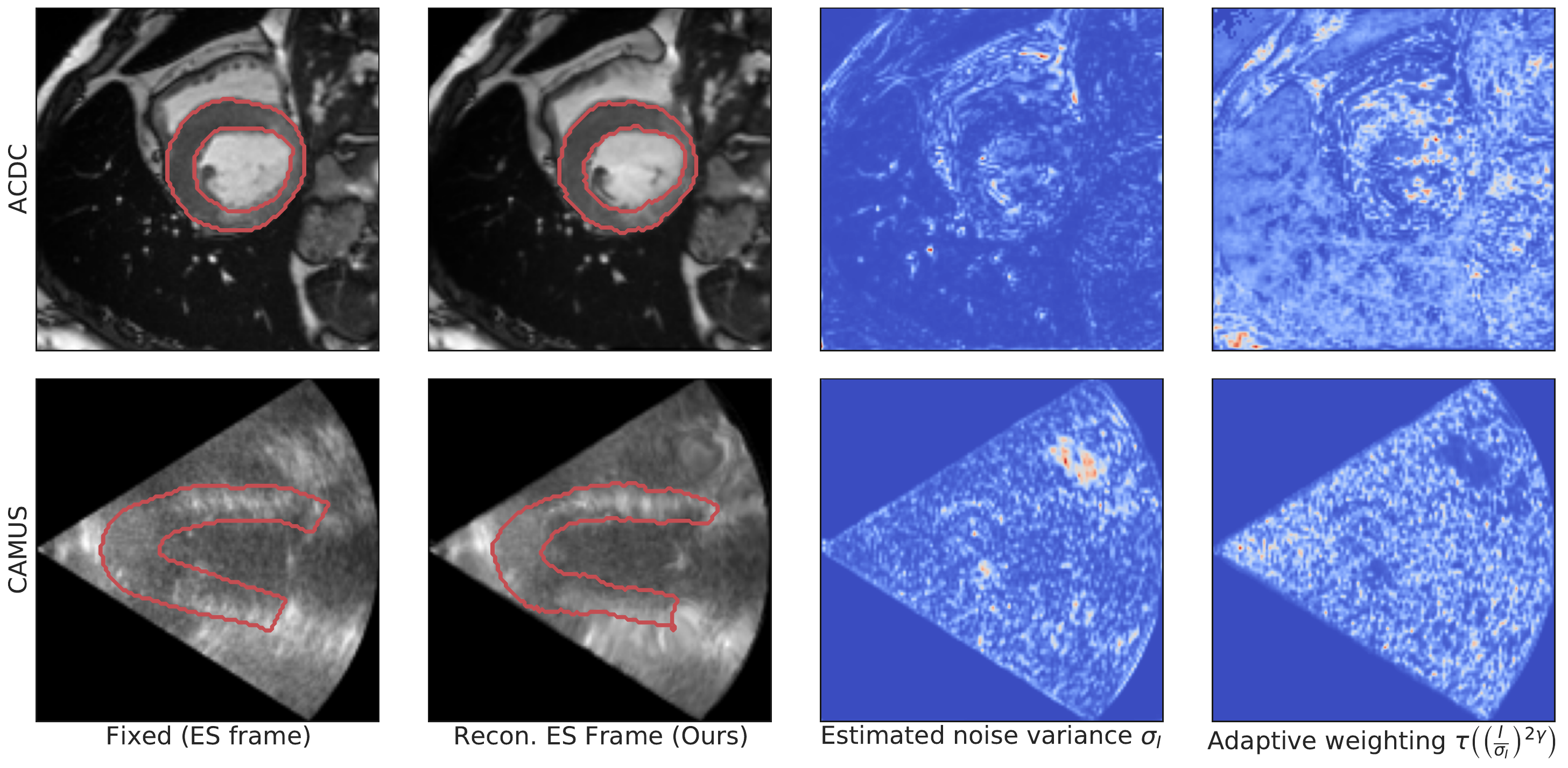}} & \includegraphics[scale=0.17]{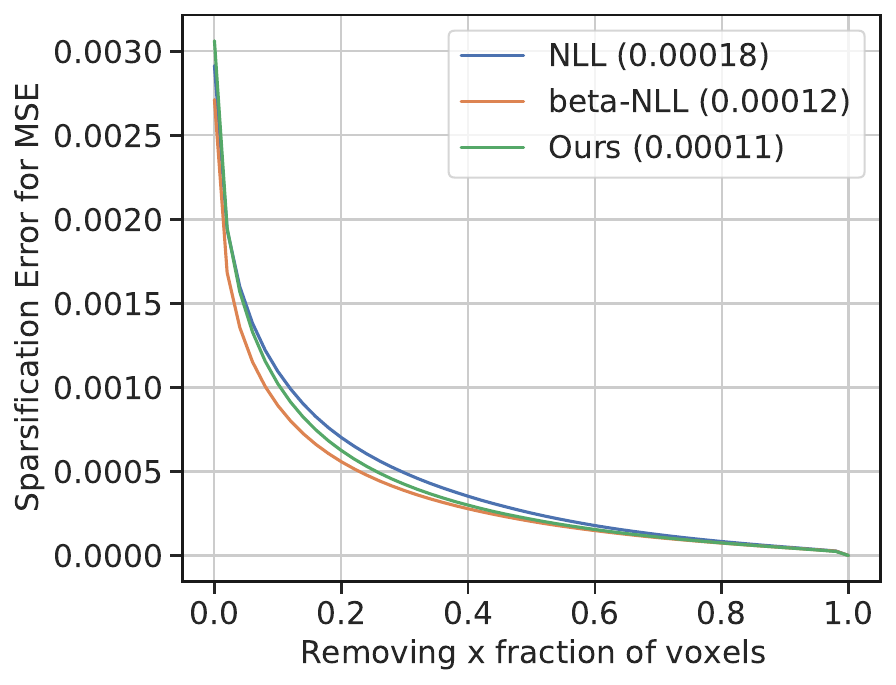} \\
      & \includegraphics[scale=0.17]{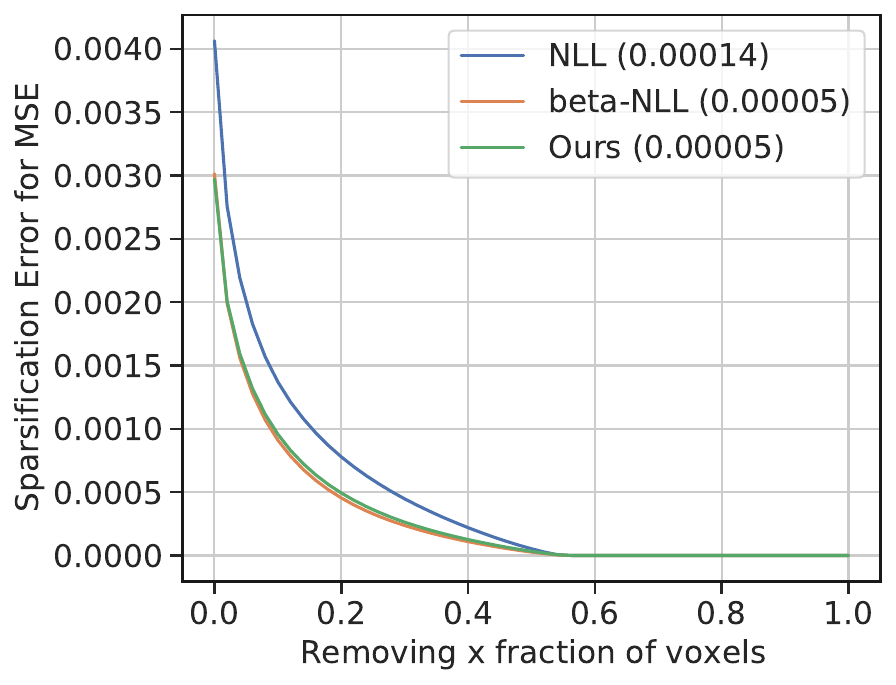}
    \end{tabular}
    \caption{Left: Estimated $\hat{\sigma}_I^2$ and the corresponding weighting map of (top row: ACDC \cite{bernard_deep_2018}; bottom row: CAMUS \cite{leclerc_deep_2019}). Right: Sparsification error plots of $\hat{\sigma}_I^2$. Both plots are from our proposed framework under Voxelmorph architecture \cite{balakrishnan_voxelmorph_2019}}
    \label{fig:ause-logsigma_image}
\end{figure}

%To quantitatively evaluate the accuracy of our estimated variance $\hat{\sigma}_I^2$, we utilized the sparsification error plots \cite{poggi_uncertainty_2020} 
\change{We utilize sparsification error plots \cite{poggi_uncertainty_2020} to provide quantitative measure of accuracy for our estimated variance $\hat{\sigma}_I^2$.} %which provide numerical measures of accuracy via the area under the error plot. 
To obtain the plots, we removed one pixel at a time from largest to smallest uncertainty magnitudes, and measured the MSE of the remaining pixels ($\mathbb{E}_{\Omega-\Omega_x}[I_f-\hat{I}_f]^2$) \cite{truong_learning_2021}. An ideal sparsification error plot should monotonically decrease, which would indicate that the estimated uncertainty map is able to correctly identify pixels with the largest errors. \Cref{fig:ause-logsigma_image} shows  our estimated uncertainty is sensible across different datasets -- based on the overall shape of the plots and the Area Under the Sparsification Error (AUSE) metrics -- with similar calibration levels compared to $\beta$-NLL \cite{seitzer_pitfalls_2022} and better calibration than NLL \cite{kendall_what_2017}. This illustrates the effectiveness of our variance estimator and optimization framework. %We additionally tested on an alternative heteroscedastic Laplacian assumption shown in the supplementary.

We further provide a qualitative visualization on both datasets, where we demonstrate that our estimated $\hat{\sigma}_I$ provides a sensible heteroscedastic noise variance map between the fixed image and reconstructed image according to our assumption (\Cref{fig:ause-logsigma_image}). From the figure, our estimated noise variance shown in the third column reflects the intensity mismatches of the corresponding regions between the fixed image $I_f$ in the first column and our reconstructed/moved image ($\hat{I}_f$) in the second column. This corroborates the validity of our heteroscedastic noise assumption (\Cref{eq:sigma_I_noise}) and shows the effectiveness of our proposed variance estimator. Furthermore, our computed adaptive weighting maps (last column, \Cref{fig:ause-logsigma_image}) accurately reflect the relative importance based on SNR, as evidenced by the highlighting of regions with potential imaging artifacts. Nonetheless, we do have failure cases, as shown in the Supp. Mat., where the myocardium (typically challenging due to irregular volumes) is considerably thin.

% leading the displacement estimator to better performance

% \subsection{Effect of $\gamma$} \label{sec:gamma}
% \input{tables/gamma_table}

% As previously mentioned in \Cref{sec:loss_function}, the exponentiated parameter $\gamma$ provides the flexibility to adjust the estimated $\hat{\sigma}_I^2$'s degree of influence on the displacement estimator during training. This can be viewed as a confidence trade-off in the current uncertainty estimate with $\gamma=0$ showing no confidence in current uncertainty estimation, reducing the adaptive weighting map to a uniform scalar map, and $\gamma=1$ indicating full belief. Considering the overall  performance across all datasets, our method seems to perform best at $\gamma=0.5$ as shown in \Cref{tab:gamma}.

%% file: tables/combined_table.tex
\begin{table}[tb]
    % \caption{\revise{Contour-based metrics compared against baselines. Units: DSC (\%) HD (vx) ASD (vx). Our method consistently improves on registration accuracy across different architectures and datasets.}}
    %\centering
    \begin{minipage}[t]{0.59\textwidth}
    % \begin{table}[tb]
    \captionof{table}{Contour-based metrics compared against baselines. Units: DSC (\%) HD (px) ASD (px). Our method consistently improves across different architectures and datasets. The second-best method is highlighted with $\dagger$.}
    \label{tab:contour}
    % \caption{\revise{Contour-based metrics compared against baselines. Units: DSC (\%) HD (vx) ASD (vx). Our method consistently improves on registration accuracy across different architectures and datasets.}}
    \centering
        \begin{adjustbox}{scale=0.81}
\begin{tabular}{llcccccc}
    \toprule
        &  & \multicolumn{3}{c}{ACDC \cite{bernard_deep_2018}} & \multicolumn{3}{c}{CAMUS \cite{leclerc_deep_2019}}  \\
      \cmidrule(lr){3-5} \cmidrule(lr){6-8} 
        &   & DSC $\uparrow$ & HD $\downarrow$ & ASD $\downarrow$ & DSC $\uparrow$ & HD $\downarrow$ & ASD $\downarrow$  \\\midrule 
        %\multicolumn{12}{c}{Gaussian additive noise assumption $p()$}\\
    \midrule
    & Undeformed & 47.98 & 7.91 & 2.32 & 66.77 & 10.87 & 2.61 \\ \midrule
    & Elastix \cite{klein_elastix_2010} & 77.26 & 4.95 & 1.28 & 80.18 & 10.02 & 1.81 \\\midrule
    \parbox[t]{2mm}{\multirow{8}{*}{\rotatebox[origin=c]{90}{CNN}}}
        & vxm (NCC) \cite{balakrishnan_voxelmorph_2019}    & 78.55 & 4.94 & 1.29 & 77.01 & 10.23 & 1.89  \\
        & vxm (MI) \cite{balakrishnan_voxelmorph_2019}    & 78.04 & 5.25 & 1.35 & 78.18 & 9.83 & 1.99  \\
        & vxm (MSE) \cite{balakrishnan_voxelmorph_2019} $\dagger$     & 80.20 & 4.64 & 1.24 & 81.76 & 8.93 & 1.70  \\
        & NLL \cite{kendall_what_2017}                      & 76.49 & 5.46 & 1.45 & 75.24 & 11.05 & 2.20  \\
        & $\beta$-NLL \cite{seitzer_pitfalls_2022}          & 78.74 & 5.07 & 1.33 & 79.75 &  9.39 & 1.93  \\
        & AdaFrame \cite{wong_adaptive_2021} & 66.38 & 5.80 & 1.67 & 77.88 & 10.54 & 1.93  \\
        & AdaReg \cite{wong_bilateral_2019} & 78.75 & 5.13 & 1.33 & 79.31 & 9.78 & 1.88  \\
        & Ours         & \textbf{80.73} & \textbf{4.57} & \textbf{1.21} & \textbf{81.96} & \textbf{8.80} & \textbf{1.66}  \\

    \midrule
    \parbox[t]{2mm}{\multirow{8}{*}{\rotatebox[origin=c]{90}{Transformer}}}
        & tsm (NCC) \cite{chen_transmorph_2022}     & 73.77 & 6.64 & 1.12 & 73.03 & 11.87 & \textbf{1.70}  \\
        & tsm (MI) \cite{chen_transmorph_2022}     & 73.57 & 6.57 & \textbf{1.11} & 74.83 & 11.94 & 1.83  \\
        & tsm (MSE) \cite{chen_transmorph_2022} $\dagger$    & 76.94 & 5.51 & 1.30 & 79.24 & 10.30 & 1.79  \\
        & NLL \cite{kendall_what_2017}                      & 73.12 & 7.22 & 1.27 & 75.08 & 11.60 & 1.79  \\
        & $\beta$-NLL \cite{seitzer_pitfalls_2022}          & 75.74 & 6.12 & 1.29 & 77.39 & 10.99 & 1.86  \\
        & AdaFrame \cite{wong_adaptive_2021} & 67.95 & 5.72 & 1.59 & 78.06 & 9.86 & 1.91  \\
        & AdaReg \cite{wong_bilateral_2019} & 76.22 & 5.68 & 1.29 & 78.12 & 10.62 & 1.84  \\
        & Ours         & \textbf{78.12} & \textbf{5.04} & 1.26 & \textbf{80.38} & \textbf{9.86} & 1.72 \\

    \bottomrule
    \end{tabular}
        \end{adjustbox}
        % \end{table}
    \end{minipage}
    \hfill
    \begin{minipage}[t]{0.4\textwidth}
    \centering
    \captionof{table}{Paired t-tests of ours v.s. $\dagger$ in \Cref{tab:contour} in terms of DSC.}
    \label{tab:p_values}
    \begin{adjustbox}{scale=0.7}
    \begin{tabular}{lcc}
    \toprule 
    & Voxelmorph  & Transmorph  \\\midrule%\midrule
    ACDC  & $p=0.02$ & $p=1.02\times 10^{-41}$ \\
    CAMUS  & $p=1.36\times 10^{-6}$ &$p=0.005$ \\
    \bottomrule
    \end{tabular}
    \end{adjustbox}
    \vspace{0.25cm}
    \captionof{table}{Quantitative evaluation on 3D private Echo dataset using vxm-based architecture.}
    \label{tab:contour_echo}
    \vspace{-0.37cm}
    \begin{adjustbox}{scale=0.83}
    \begin{tabular}{lccccccccc}
    \toprule
        &  \multicolumn{3}{c}{Private 3D Echo} \\
      \cmidrule(lr){2-4} 
        &  DSC $\uparrow$ & HD $\downarrow$ & ASD $\downarrow$ \\\midrule 
        %\multicolumn{12}{c}{Gaussian additive noise assumption $p()$}\\
    \midrule
        Voxelmorph \cite{balakrishnan_voxelmorph_2019}    & 74.61 & 5.59 & 0.94 \\
    NLL \cite{kendall_what_2017}                      & 72.59 & 6.35 & 0.97 \\
        $\beta$-NLL \cite{seitzer_pitfalls_2022}          & 73.81 & 5.77 & 0.96 \\
        AdaFrame \cite{wong_adaptive_2021} & 73.01 & 5.74 & 0.94 \\
        AdaReg \cite{wong_bilateral_2019} & 73.41 & 5.91 & 0.95 \\
        Ours          & \textbf{75.04} & \textbf{5.55} & \textbf{0.93} \\
    \bottomrule
    \end{tabular}
    \end{adjustbox}
    \vspace{0.25cm}
    
    \captionof{table}{Effect of $\gamma$ in \Cref{eq:displacement_loss} for $\hat{\sigma}_I^2$ estimation.}%Although our method is not extremely sensitive to the choice of $\gamma$, we can still observe that $\gamma=0.5$ provides the best overall performance.}
    \label{tab:gamma}
    \vspace{-0.37cm}
    \begin{adjustbox}{scale=0.57}
    % \begin{subtable}{0.4\linewidth}
        \begin{tabular}{lccccccccc}
    \toprule
        & \multicolumn{3}{c}{ACDC \cite{bernard_deep_2018}} & \multicolumn{3}{c}{CAMUS \cite{leclerc_deep_2019}}  \\
   \cmidrule(lr){2-4} \cmidrule(lr){5-7} 
         & DSC $\uparrow$ & HD $\downarrow$ & ASD $\downarrow$ & DSC $\uparrow$ & HD $\downarrow$ & ASD $\downarrow$  \\
    \midrule
        Ours ($\gamma=0.25$)  & 79.74 & 4.74 & 1.26 & \textbf{82.07} & 8.53 & \textbf{1.65}  \\
        Ours ($\gamma=0.5$)   & \textbf{80.73} & \textbf{4.57} & \textbf{1.21} & 81.96 & 8.80 & 1.66  \\
        Ours ($\gamma=0.75$)  & 80.00 & 4.69 & 1.24 & 81.82 & \textbf{8.45} & 1.66  \\
        Ours ($\gamma=1$)     & 79.78 & 4.71 & 1.25 & 81.31 & 9.08 & 1.69  \\
    \bottomrule
    \end{tabular}
    \end{adjustbox}%
    % \end{table}
    % \begin{table}[tb]
    % \caption{a test}
    
    % \end{table}
    \end{minipage}
\end{table}

%% file: sections/5_conclusion.tex
\section{Conclusion}
\label{sec:conclusion}
We propose a heteroscedastic uncertainty estimation framework for probabilistic unsupervised image registration to adaptively weight the displacement estimation with \textit{relative} $\gamma$-exponentiated signal-to-noise, which improves registration performance from the commonly used homoscedastic assumption while also providing accurate and sensible uncertainty measures. Our proposed framework consists of a displacement and a variance estimator, optimized under an alternating collaborative strategy. We demonstrate the effectiveness and versatility of our proposed framework on two representative registration architectures across diverse cardiac datasets and show consistent statistically significant improvements over other baselines. Though our proposed framework is promising, it still relies on a manually crafted adaptive map on the data fidelity term, which might not be able to fully reflect the complicated characteristics of large-scale real-world data. Future work will aim to explore a more data-driven objective with further validation of clinical datasets for more potential impact.

% \section{Acknowledgment}
% This work is supported by NIH grant R01HL121226.

%% file: sections/supplementary.tex
\clearpage
\setcounter{page}{1}
\setcounter{section}{0}
% \begin{center}
%     \Large \textbf{Supplementary Materials}
% \end{center}
\renewcommand{\thesection}{\Alph{section}} 
\paragraph{\textbf{Dataset and implementation details}}%\hfill\break
% \section{Dataset and implementation details}
\begin{itemize}
    \item \textbf{ACDC (2D MRI) \cite{bernard_deep_2018}}.
    We randomly selected 80 patients for training, 20 for validation, and 50 for testing. ED and ES image pairs are extracted from each sequence in a slice-by-slice manner from the longitudinal stacks. We center crop each slice pair to $128\times 128$ w.r.t. myocardium centroid, yielding 751 image pairs for training, 200 for validation, and another 538 for testing.
    \item \textbf{CAMUS (2D Echo) \cite{leclerc_deep_2019}}.
    We resize each image pair to size $128\times 128$ and randomly select 300 subjects for training, 100 subjects for validation, and 100 subjects for testing. This yields in total 600 image pairs for training, 200 pairs for validation, and 200 pairs for testing.
    \item \textbf{Private 3D Echo}.
    The private 3D echo dataset contains 99 cardiac ultrasound scans. ED and ES frames are manually identified and myocardium segmentation labels are provided for each sequence by experienced radiologists. Each 3D image is resized to $64\times 64\times 64$. We randomly select 60 3D pairs for training, 19 pairs for validation, and another 20 pairs for testing.
    \item \textbf{Implementation details}.
    All our experiments are conducted under the Pytorch framework and trained on NVIDIA V100/A5000 GPUs. The architecture of the variance estimator is implemented based on a U-Net. We use $\lambda=0.01$ as the hyperparameter in \Cref{eq:displacement_loss}. Both displacement and variance estimators are trained with learning rates $1 \times 10^{-4}$ for 300 epochs.
\end{itemize}

\paragraph{\textbf{Incorporating displacement uncertainty.}}
% \section{Incorporating displacement uncertainty}
To further demonstrate the versatility, we conducted a direct extension by simultaneously estimating heteroscedastic displacement uncertainty with the isotropic assumption. We add an additional layer in the displacement estimator to predict $\hat{\sigma}_z$, where $\hat{\sigma}_z(x)\in\mathbb{R}$ and the original prediction as displacement mean $\hat{\mu}_z$. We train our proposed displacement estimator using objective:

$\mathcal{L}_{\theta} = \mathbb{E}_\Omega \left[ \mathcal{T}\left[ \left(\frac{I_f}{\floor{\hat{\sigma}_I}}\right)^{2\gamma}\right] [I_f-\hat{I}_f]^2 + \alpha\left(\hat{\sigma}_z^2-\log \hat{\sigma}_z^2\right) + \lambda \|\nabla \hat{z}\|^2 \right]$ derived from \Cref{eq:full_exp}, with $\hat{z}$ sampled from distribution $\hat{z}\sim\mathcal{N}(\hat{\mu}_z,\hat{\sigma}_z^2\mathbb{I})$ during training with re-parameterization trick. %practice, we utilize the re-parameterization trick to avoid interruption in gradient-based optimization when doing the above sampling. 
We compare the quality of our predicted displacement along with its uncertainty estimate $\hat{\sigma}_z^2$ with vxm-diff \cite{dalca_unsupervised_2019}. 
% \label{sec:exp_sigma_z}
% \vspace{-0.7cm}
\begin{figure}[h!]
    % \centering
    \begin{minipage}{0.5\linewidth}
    %\begin{flushleft}
    \centering
    \includegraphics[scale=0.15]{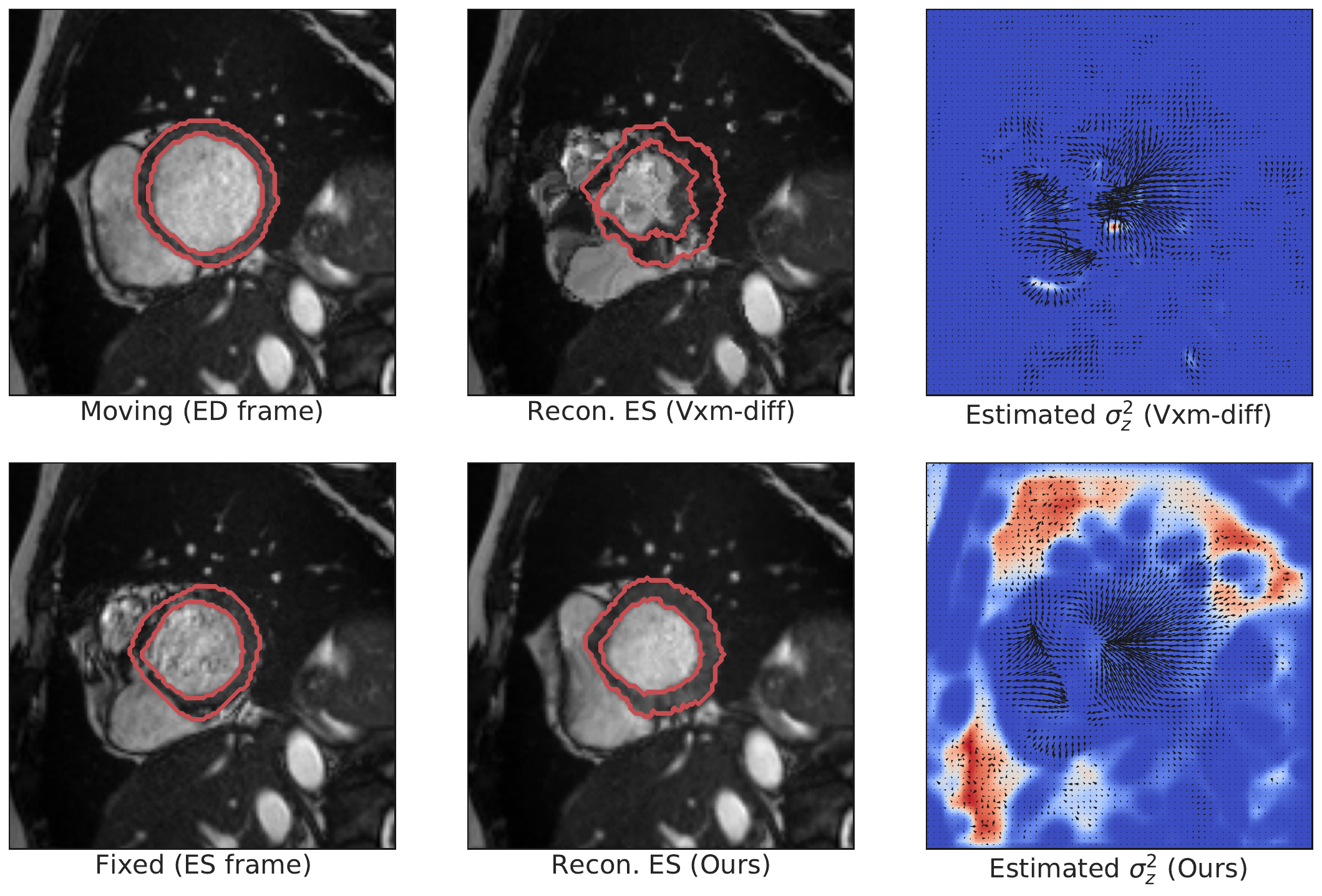} 
    \caption{Comparison of $\hat{\sigma}_z^2$ between our vxm-based framework and vxm-diff \cite{dalca_unsupervised_2019}.}
    \label{fig:logsigma_flow}
        %\end{flushleft}
    \end{minipage}%\hfill
    \begin{minipage}{0.44\linewidth}
    \centering
    \captionof{table}{Our method raises the upper bound on registration accuracy while providing useful displacement uncertainty estimates $\hat{\sigma}_z$.}
    \label{tab:sigma_z}
    %\begin{flushright}
    \begin{adjustbox}{scale=0.53}
    \begin{tabular}{lcccccccc}
    \toprule
        & \multicolumn{2}{c}{Uncertainty} & \multicolumn{3}{c}{ACDC \cite{bernard_deep_2018}} & \multicolumn{3}{c}{CAMUS \cite{leclerc_deep_2019}}  \\
    \cmidrule(lr){2-3}  \cmidrule(lr){4-6} \cmidrule(lr){7-9} 
        & $\sigma_z^2$ & $\sigma_I^2$  & DSC $\uparrow$ & HD $\downarrow$ & ASD $\downarrow$ & DSC $\uparrow$ & HD $\downarrow$ & ASD $\downarrow$  \\\midrule 
        %\multicolumn{12}{c}{Gaussian additive noise assumption $p()$}\\
    \midrule
        Vxm \cite{balakrishnan_voxelmorph_2019}   & \xmark &\xmark  & 80.20 & 4.64 & 1.24 & 81.76 & 8.93 & 1.70  \\\midrule
        Vxm-diff \cite{dalca_unsupervised_2019}   & \cmark & \xmark & 76.19 & 5.75 & \textbf{1.19} & 76.74 & 10.76 & 1.88  \\
        Ours         & \cmark & \xmark & 79.80 & 4.74 & 1.22 & 81.47 &  8.67 & 1.69  \\\midrule
        Ours         & \xmark & \cmark & \textbf{80.73} & \textbf{4.57} & 1.21 & \textbf{81.96} & 8.80 & 1.66  \\
        Ours         & \cmark & \cmark & 79.87 & 4.62 & 1.20 & 81.91 & \textbf{8.54} & \textbf{1.65}  \\
    \bottomrule
    \end{tabular}
    \end{adjustbox}

        %\end{flushright}
    \end{minipage}
\end{figure}

We present our quantitative results in \Cref{tab:sigma_z}, illustrating the superiority of our formulation. We further present the qualitative visualization as shown in \Cref{fig:logsigma_flow}, demonstrating that our estimated heteroscedastic uncertainty $\hat{\sigma}_z^2$ accurately captures the randomness in the displacement prediction more accurately. %due to factors such as motion ambiguity. %Additionally, we note that both our formulation and Voxelmorph as shown in \cref{tab:sigma_z} fail to improve registration performance by incorporating displacement uncertainty. We argue that by incorporating displacement uncertainty in both formulations does not contribute to learning a more accurate correspondence but rather provides an uncertainty estimate with the cost of accuracy, which results from the randomness in the sampling process in both frameworks ($\hat{z}\sim\mathcal{N}(\hat{\mu}_z,\hat{\sigma}_z^2\mathbb{I})$) also degrades the performance. %On the contrary, our proposed adaptive weighting scheme based on relative signal-to-noise not only provides additional uncertainty estimates but also improves registration performance. 

\paragraph{\textbf{Additional private 3D Echo results.}}%\hfill\break

\begin{figure*}[tb]
    \centering
    \begin{tabular}{c}
         \includegraphics[scale=0.3]{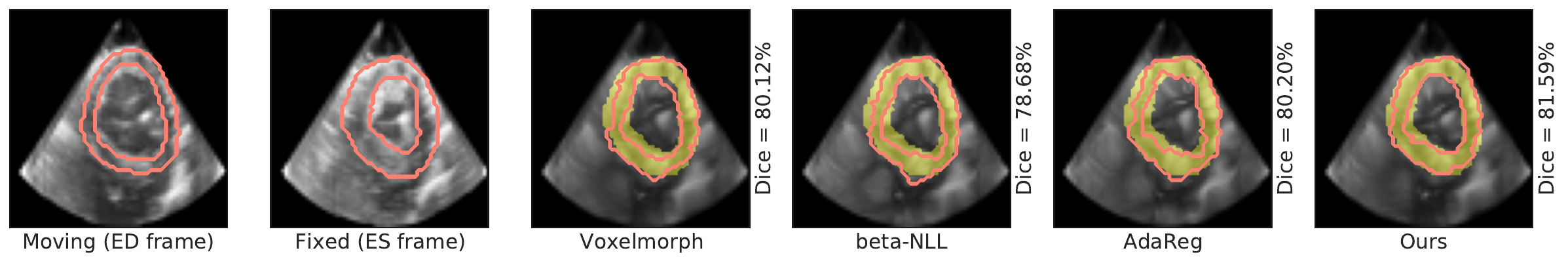} %\midrule
    \end{tabular}
    
    \caption{Qualitative evaluation for our private 3D Echo dataset on voxelmorph architecture. We extract cross-sectional slices from the 3D volume for visualization. We overlay ground truth segmentation in yellow for comparison.}
    \label{fig:contour_echo}
\end{figure*}

\begin{figure}[tb]
    \centering
    \begin{tabular}{c|c}
         \includegraphics[scale=0.185]{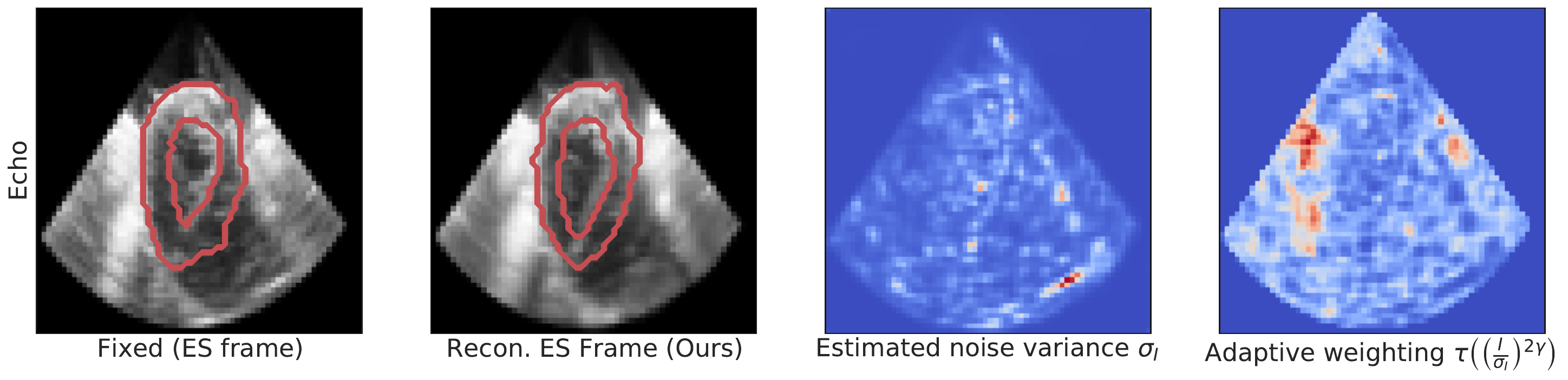} & \includegraphics[scale=0.185]{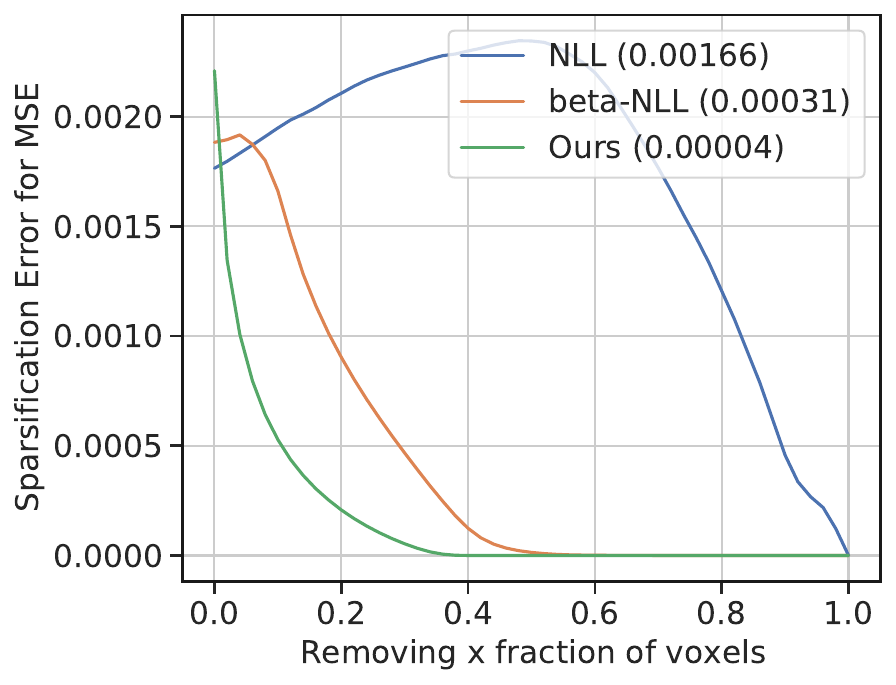}
    \end{tabular}
    \caption{Left: Estimated $\hat{\sigma}_I^2$ and the corresponding weighting map of our proposed framework under Voxelmorph architecture \cite{balakrishnan_voxelmorph_2019} using our private 3D Echo dataset. Right: Sparsification error plots of $\log\hat{\sigma}_I^2$ on our private 3D Echo dataset.}
    \label{fig:logsigma_image_echo-ause_echo}
\end{figure}

\noindent We present our qualitative result in \Cref{fig:contour_echo} for registration accuracy and left \Cref{fig:logsigma_image_echo-ause_echo} for noise heteroscedastic variance evaluation. We also quantitatively evaluate the result by repeating the sparsification error plot similar to \Cref{sec:exp_sigma_I}. We observe that our predicted $\hat{\sigma}_I$ achieves a better error curve than $\beta$-NLL and NLL, which is consistent with our main results shown in \Cref{fig:ause-logsigma_image}.%We present the quantitative evaluation with \cref{tab:contour_echo}. We note that our proposed framework outperforms other baselines under the Voxelmorph architecture while achieving comparable performance with the vanilla version under the Transmorph architecture. This might be due to the size of our private 3D testing set being small with only 20 cases, which we will further evaluate on larger 3D \textit{in vivo} animal datasets. 
% \vspace{-0.7cm}

\begin{figure}[h!]
    \centering
    \includegraphics[scale=0.3]{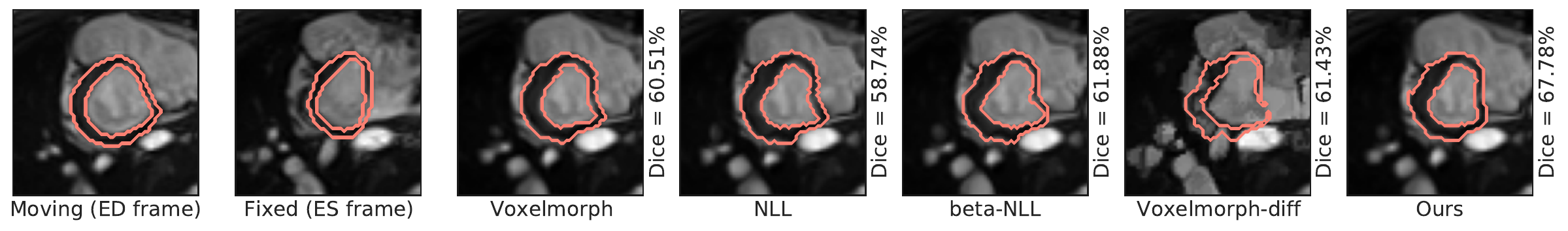}
    \caption{An example of failure case on ACDC dataset.}
    \label{fig:ACDC_failure}
\end{figure}

\paragraph{\textbf{Failure case.}}
We present an example shown in \Cref{fig:ACDC_failure} that all methods fail to match myocardium when it is considerably thin.
